\documentstyle[prl,aps,psfig,twocolumn]{revtex} 

\begin{document}
\draft
\twocolumn[\hsize\textwidth\columnwidth\hsize\csname
@twocolumnfalse\endcsname
\title{Cold and Warm Denaturation of Hydrophobic
Polymers.}
\author{Paolo De Los Rios$^1$, Guido Caldarelli$^2$}
\address{$^1$Institut de Physique Th\'eorique, 
Universit\'e de Fribourg, CH-1700, Fribourg, Switzerland.}
\address{$^2$INFM Sezione di Roma1, Dip. Fisica Universit\`a La Sapienza, 
Ple A. Moro 2, 00185 Roma Italy.}
\date{\today}
\maketitle

\begin{abstract}
We introduce a polymer model where the transition 
from swollen to compact configurations is due 
to interactions between the monomers and the solvent.
These interactions are the origin of the effective
attractive interactions between hydrophobic amminoacids
in proteins.
We find that in the low and high temperature phases polymers
are swollen, and there is an intermediate phase where the
most favorable configurations are compact. 
We argue that such a model captures in a single framework
both the cold and the warm denaturation experimentally
detected for proteins. Some consequences for protein folding are discussed.
\end{abstract}
]
\narrowtext
 
Modeling polymers and polymer collapse, beyond being a 
challenge of great theoretical interest, is extremely important
for many different applications~\cite{Vanderzande}. In particular, 
in connection to the protein folding problem, the collapse of polymers
has gained a special status in statistical physics. 
The simplest models of proteins describe them as polymers
whose building monomers (amminoacids) can be of two different types, either
hydrophobic (H) or polar (P) (such a simplification
was introduced in the so-called HP model \cite{LD89}). 
The protein folds through attractive H-H interactions.
(Homo)Polymers with attractive monomer-monomer interactions
are known to undergo a phase transition from a high-temperature swollen phase
to a low-temperature compact one. 
The presence of attractive interactions between 
hydrophobic amminoacids justifies then the collapse of the protein 
below the folding temperature.
In particular, trying to maximize the number of H-H contacts, most 
hydrophobic monomers will be found in the core of the protein, in agreement 
with experimental observations\cite{Creighton}.

It is important to understand the origin of the
attractive H-H interaction: as their name suggests, hydrophobic amminoacids
do not like being in contact with water. As a consequence, hydrophobic 
amminoacids in solution tend to aggregate in order to minimize
the area exposed to water. This tendency to aggregate can be modeled
as an attractive interaction. In proteins, hydrophobic amminoacids try to hide
away from water, burying themselves in the core of the protein (whose surface
is mainly made of polar, hydrophilic amminoacids).

At temperatures much lower than the folding transition, the HP model
(and models derived from it) predicts that proteins 
stay in a compact state, with
a greater and greater probability to be in their ground state.
This principle has also been used to classify different amminoacid
sequences as good or bad folders (from a thermodynamical point of view)
according to {\it e.g.} the gap between the ground state and the first excited states,
or the uniqueness of the ground state on a compact configuration\cite{SVMB96}.

Recently, nonetheless, there has been a growing evidence for the so-called
{\it cold denaturation} of proteins: at low temperatures 
proteins such as $\beta$-Lactoglobulin A, myoglobulin, apomyoglobulin, Ribonuclease A,
and {\it Escherichia Coli}'s HPr~\cite{biochimica}
unfold to a swollen configuration~\cite{low temp}. Cold denaturation
is assumed to be a general property of all globular proteins\cite{Privalov90}. 
These experimental findings are clearly incompatible with 
the predictions of the HP model.

In this Letter we propose a mechanism
driving the collapse of a polymer, mimicking the
nature of the hydrophobic effect. Furthermore, we find within a single
framework both a cold and a warm collapse transitions, in agreement
with experimental observations for hydrophobic amminoacids.
The model is built on the present understanding of the microscopic
organization of the water/hydrophobic molecules system.

Hydrophobic molecules are essentially
non-polar entities, weakly, if at all, interacting with water
molecules. Therefore, there is no true repulsion 
between water and hydrophobic molecules. 
We can get an insight in the energetics of 
water/hydrophobic molecules systems from experiments.
In the case of pentane\cite{PG88} (hydrocarbons
are in general hydrophobic), one finds an enthalpy of transfer 
$\Delta H_{a\to s}$ from the aggregate phase to the
aqueous solution phase that is strongly temperature dependent: at low
temperatures $\Delta H_{a\to s} <0$, and at high temperatures $\Delta H_{a\to s} >0$,
showing that at low temperatures the solute phase is {\it energetically}
more favourable than the aggregate one; at high temperatures the situation
reverses.

Microscopically, a hydrophobic molecule in solution
displaces water molecules that, at a low enough temperature, 
build an ice-like cage around the hydrophobic molecule, 
giving origin to a structure that is energetically more favorable than bulk, 
liquid water\cite{Creighton}. This energetic
gain gives the major contribution to the minimization of the free energy,
and as a consequence the system tries to maximize
the number of cages by exposing as many hydrophobic molecules as possible,
hence the cold denaturation of proteins and $\Delta H_{a\to s}<0$. 
At higher temperatures
water around hydrophobic molecules can no more
form the cages, but due to the steric constraints imposed by the molecule,
they cannot even fully exploit the energetically favorable hydrogen bonding of
bulk, liquid water. As a result, disordered water molecules around
hydrophobic amminoacids are energetically less favorable than bulk water 
($\Delta H_{a\to s}>0$): 
proteins try therefore to hide their hydrophobic parts in their core.
This is what is usually referred to as the {\it hydrophobic effect},
and it is the consequence of a complex interplay between energy
minimization and entropy maximization.  
Unfortunately, the detailed behavior of water at low temperatures
is still far from being well understood from a 
microscopic point of view; the nature
of the hydrogen bond itself is still a matter of controversy and of deep 
investigations\cite{ISPHBT99}. Therefore it is still 
not possible to give a detailed description of the hydrophobic
amminoacid-water organization and of the energies associated to 
the cages and to the disordered configurations.


We model the polymer as a self-avoiding walk (SAW) on a lattice.
In every lattice site (except those occupied by the polymer) there is 
a Potts variable with $q$ states (labelled for convenience from $0$
to $q-1$), representing a group of water molecules in $q$
different collective states. We associate the state $q=0$ to the
cage configuration, energetically favorable when
water is in contact with the polymer, and the
remaining $q-1$ states to disordered, unfavorable, configurations.
The Hamiltonian of the system is
\begin{equation}
H = \sum_{i=1}^N \sum_{j n.n. i}{}' (-J \delta_{s_j,0} + K (1-\delta_{s_j,0}))\;.
\label{hamiltoniana}
\end{equation}
The first sum runs over the $N$ monomers of the polymer. 
The second primed sum runs over the water occupied
nearest neighbors of each monomer. The interaction constants $J$ and $K$
(both positive) represent respectively the energies of the cage configurations
and of the disordered ones with respect to bulk water.
There are no monomer-monomer interactions.

Some kind of water-protein interaction with a similar description
of the water degrees of freedom was already proposed in~\cite{HJSZ97}
with the same motivations. Yet, in that model, the water-protein interaction
was introduced in a ``mean field" fashion, and moreover protein folding
was described by collective hierarchical variables hiding the
microscopic description, that we believe fundamental to
give qualitative, but also quantitative, predictions.

Starting from (\ref{hamiltoniana}) we can write the partition
function of the system as 
\begin{equation}
Z_N = \sum_{C} Z_N(C)
\label{partition}
\end{equation}
where $Z_N(C)$ is the partition function associated to a single
configuration $C$. It is important to
observe that the maximum number of water sites in contact
with the polymer is
$M=2(d-1)N+2$ for a hypercubic lattice in $d$ dimensions. 
For general polymer configurations 
the number of contacts is smaller than $M$. Nonetheless, all of these
$M$  water sites must be taken into account in order to give
the correct weight to all the $Z_N(C)$'s. Moreover, the way 
Hamiltonian (\ref{hamiltoniana}) has been written implies
that a single water site can be counted more than once.
It is then possible to write a fairly simple expression for
the configuration partition function:
\begin{equation}
Z_N(C) = q^{n_0(C)} \prod_{l=1}^{z} Y_l(C)
\label{conf part func}
\end{equation}
with
\begin{equation}
Y_l(C) = \left((q-1) e^{-\beta l K } + e^{\beta l J} \right)^{n_l(C)}
\label{factor}
\end{equation}
where $n_l(C)$ is the number of water sites with $l$ polymer contacts
and $z$ is the coordination number of the lattice. As usual, $\beta = 1/k_B T$
and we take $k_B =1$.
Analogously, it is possible to write similar expressions (even though a little
more complicated) for the internal energy $U(C)$ of a configuration
(and then calculate the internal energy $U = \sum_C U(C)$). 

We analyze the thermodynamic behavior of the system in $d=2$
by means of exact enumeration techniques on the Manhattan lattice
for polymers of length up to $N=25$ monomers.
The Manhattan lattice is a two-dimensional lattice on which 
rows (columns) are alternately left/right (up/down) oriented.
The physics of polymer collapse in the well-known case of 
attractive monomer-monomer interactions is the same on the Manhattan lattice
as on regular two-dimensional lattices (there is still some
debate on non-universalities in the values of some exponents
~\cite{Vanderzande,CCGP98}). We are therefore confident that
our exact enumeration on the Manhattan lattice
describes the correct behavior of the polymer on 
regular Euclidean lattices.

\begin{figure}
\centerline{\psfig{file=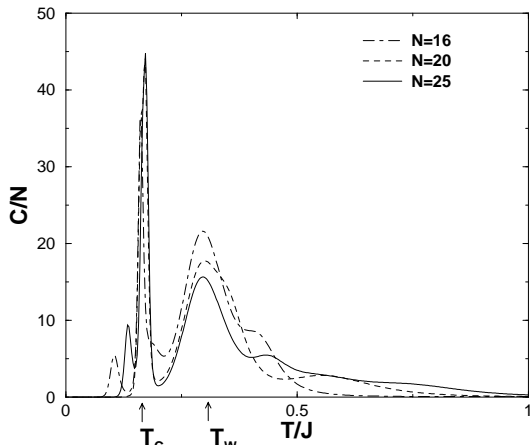,width=7.0cm,angle=0}} 
\caption{Specific heat of the system for different
polymer lengths $N=16,20,25$. Here $K/J=2$ and
$q=16000$. The cold
and warm denaturation temperatures, $T_C$ and $T_W$, are marked with arrows.} 
\label{Fig: fig1}
\end{figure}

We calculated the specific heat per monomer
of the system as the derivative of $U$
with respect to the temperature $T$, see Fig.\ref{Fig: fig1}.
We used $K/J=2$ and $q=16000$ (both $K$ and the temperature
can be normalized with respect to $J$). 
In Fig.\ref{Fig: fig1} two peaks of the specific heat appear, 
corresponding to temperatures
$T_C$ and $T_W$. By direct inspection of the values of the configuration 
partition function in eq.(\ref{conf part func}), we find that below 
$T_C$ the most probable
configurations are swollen, maximizing the number of 
water-polymer contacts. Between $T_C$ and $T_W$ the polymer collapses, 
and the most probable 
configurations are the compact ones (in the particular cases of
$N=16,25$ we verified that they correspond to square configurations). 
Finally, 
for $T > T_W$ the weight of compact and swollen configurations
becomes comparable and the warm denaturation takes place.

\begin{figure}
\centerline{\psfig{file=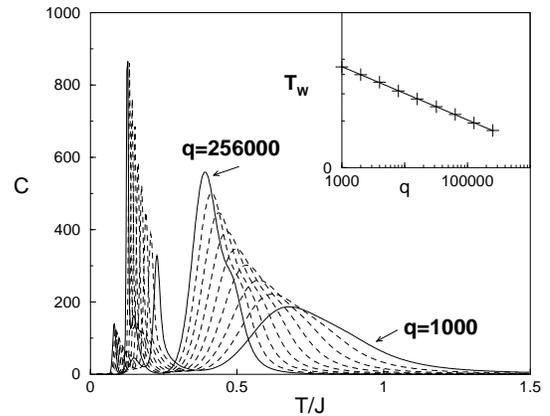,width=7.0cm,angle=0}} 
\caption{Specific heat vs. temperature for $K/J=4$ and 
different values of $q$, namely $q=2^n\cdot1000$ with $n=0,...,8$,
and $N=16$.
In the inset the cold denaturation temperatures ($+$) are 
shown vs. $q$ (they have already been shifted by the constant term $0.22$); 
the solid line is the fit $T_W - 0.22 \simeq 1.47 q^{-0.17}$} 
\label{Fig: fig2}
\end{figure}

A number of questions arise for this problem: as it can be seen 
from Fig.\ref{Fig: fig1}, the low-temperature peak in the specific heat
is stable against the length of the polymer, a clear indication of a true phase
transition. On the contrary, the high-temperature specific heat peak
seems to flatten as the length of the polymer is increased, indicating that
the high temperature denaturation transition could disappear as the length of
the polymer increases. This scenario needs further investigations.
Moreover, the behavior of the two peaks as a function of $K/J$ is such 
that they coalesce when $K/J=1$. The height of the warm denaturation
peak depends also on $q$, increasing as $q$ increases, as shown in 
Fig.\ref{Fig: fig2} (as a consequence, 
the flattening of the warm denaturation
peak depends on the interplay between $q$ and $N$).
Also the cold and warm denaturation temperatures depend on $q$. 
In the inset in Fig.\ref{Fig: fig2} we temptatively show that the
dependance of $T_W$ on $q$ follows a power-law: the best fit, quite stable
adding points on both extrema ($q$ small and large) is
$T_W \simeq 0.22 + 1.47 q^{-0.17}$.

\begin{figure}
\centerline{\psfig{file=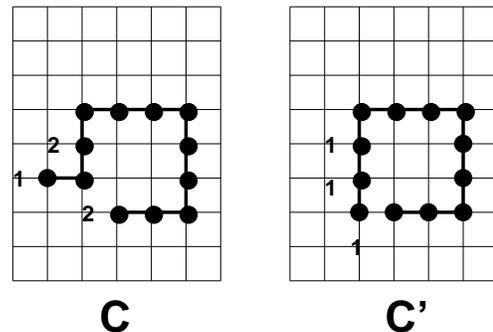,width=6.5cm,angle=0}} 
\caption{Two different configurations differing only for
a monomer-monomer contact. These configurations are used to calculate 
$\epsilon$ as in Fig.3. The water sites that changed their number
of contacts from $C$ to $C'$ are indicated with their number of contacts.} 
\label{Fig: fig3}
\end{figure}

\begin{figure}
\centerline{\psfig{file=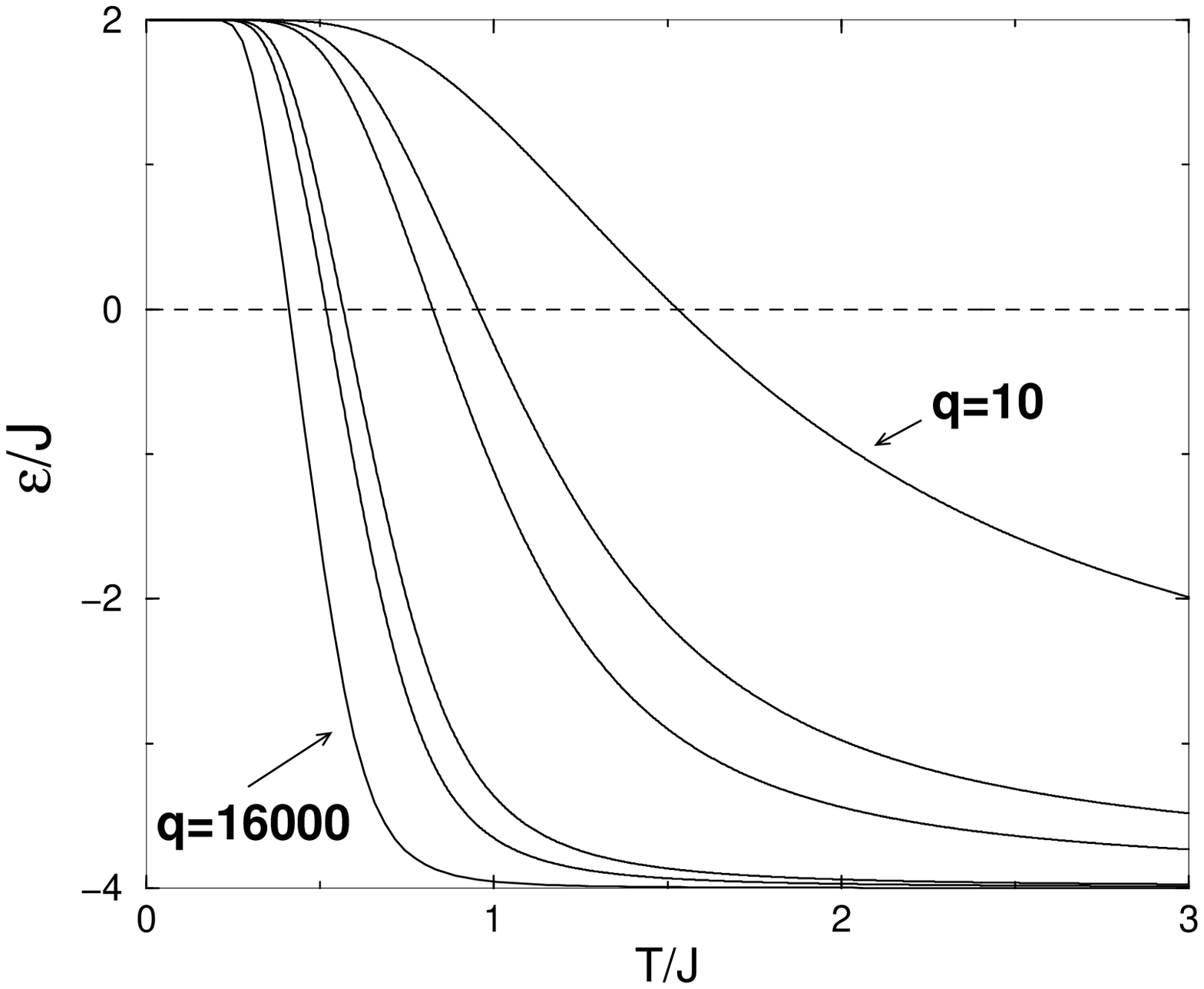,width=7.0cm,angle=0}} 
\caption{Values of the effective monomer-monomer interaction energy
as computed from (5) and the configurations in Fig.2.
The ratio $K/J=2$ and the values of $q=16000,2000,1000,100,50,10$
have been used. The transition between repulsive and attractive energies
becomes more and more abrupt as $q$ increases.} 
\label{Fig: fig4}
\end{figure}

Hydrophobic amminoacids are modeled
through an effective mutual attractive interaction. In this way
the solvent degrees of freedom can be neglected. Here we show that the
Hamiltonian in eq.(\ref{hamiltoniana}) and the corresponding
configuration partition function in eq.(\ref{conf part func})
give rise to an effective monomer-monomer attractive interaction.
Let us consider the two polymer configurations $C$ and $C'$ represented in
Fig.\ref{Fig: fig3}. Configuration $C'$ is characterized by having a
monomer-monomer contact, as opposed to configuration $C$.
If we forget about the solvent, and we attribute the difference
of the two partition functions $Z(C)$ and $Z(C')$ to a monomer-monomer
interaction energy $\epsilon$, we can write
\begin{equation}
\epsilon = - \frac{1}{\beta} \ln\frac{Z(C')}{Z(C)}
\label{effective interaction}
\end{equation}
The energy $\epsilon$ is represented in Fig.\ref{Fig: fig4} as a function
of the temperature. 
The low and high temperature limits can be easily computed as
\begin{eqnarray}
\epsilon &\simeq& 2J \hspace{3cm} T\to 0\nonumber \\
\epsilon & \simeq& -2\left(K - \frac{J+K}{q}\right) \hspace{.5cm} T\to \infty
\label{limits}
\end{eqnarray}
The effective monomer-monomer interaction energy is therefore
repulsive for small temperatures, and attractive for high temperatures,
mimicking an effective attractive H-H interaction.
Yet, through this model,
it is extremely clear that the hydrophobic attractive interaction is present
only when hydrophobic monomers are in contact with water.
We can picture a situation where only a few monomers are hydrophobic.
In this case they try to hide in the core of some globule. Within the 
core there are no more interactions between them, if not
residual ones such as Van der Waals forces. As a consequence,
in the core of a protein, the hydrophobic interaction
can not justify the maximization of the number of contacts, neither
it can discriminate between the native state of the protein
and the {\it molten globule} state, where the
hydrophobic protein core is believed to be in an amorphous state.

The absence of interactions in the core of the collapsed polymer
is the cause of the weakness of the warm denaturation transition:
for a polymer of $N$ monomers, the energy of compact configurations
grows like $N^{1/2}$, since it is just a perimeter effect. On the contrary,
we can expect an exponential number of ways to open up the compact polymer, and
therefore the entropy grows like $N$. On larger and larger length scales
the entropic term dominates and compact configurations are
not stable anymore. On the contrary, cold denaturation  
is a true transition: again the entropy grows like $N$, but in this
case the energy is extensive too.

In conclusion, we have introduced a model Hamiltonian for the collapse
of a polymer interacting with the solvent, but without 
monomer-monomer interactions. Such a model mimics (in an extremely
simplified way)
the behavior of hydrophobic polymers in water: as a result we find the presence
of both a cold and a warm denaturation temperatures, as experimentally
found for proteins (whose folding is believed to be induced
by hydrophobic interactions). Between $T_W$ and $T_C$ the polymer
collapses to compact configurations. Indeed the model tries to capture some
of the essential physics of the water-hydrophobic amminoacids
interaction at a microscopic level, and as a byproduct (necessary for
consistency), it justifies the presence of  an effective attractive
H-H interaction at high temperatures. Yet, this model rises some
questions about the reliability of using an attractive H-H interaction
also in the core of proteins (where water is absent) at odds with
models presently in use.
This model further shows that the solvent degrees of freedom can, and should, 
be taken into account in the formulation of protein models, at least in a
simplified way. In particular it calls for a much better understanding 
of the physics of water around hydrophobic amminoacids.
We are presently working on larger computer simulations of this model
for polymers embedded in lattice different from the Manhattan one.
We are also considering slight modifications of this prototype model to
take into account a a better description
of the water degrees of freedom and of their energetics.
Nevertheless, we believe to have pointed out the essential ingredients to
describe in a single framework the processes of cold and warm denaturation.

We thank P. Bruscolini, L. Casetti, A. Hansen, 
G. Tiana, Y.-C. Zhang, M. Ceppi, 
L. Pirola, F. Portis and G. Solinas for useful discussions and comments.
This work was partially supported by the European Network Contract
FMRXCT980183.

\begin{thebibliography}{99}

\bibitem{Vanderzande} C. Vanderzande, {\it Lattice Models of Polymers} (Cambridge
University Press, Cambridge 1998).

\bibitem{LD89} K.F. Lau and K.A. Dill, Macromolecules {\bf 22}, 3986 (1989).

\bibitem{Creighton} T.E. Creighton, {\it Proteins. Structures and Molecular
Properties} (W.H. Freeman \& Company, New York 1993).

\bibitem{SVMB96} F. Seno, M. Vendruscolo, A. Maritan and J.R. Banavar,
Phys. Rev. Lett. {\bf 77}, 1901 (1996).

\bibitem{biochimica} see {\it e.g.}
N.C. Pace and C. Tanford, Biochemistry {\bf 7}, 198 (1968);
Y.V. Griko and P.L. Privalov, Biochemistry {\bf 31}, 8810 (1992);
J. Zhang, X. Peng, A. Jonas and J. Jonas, Biochemistry {\bf 34}, 8631 (1995);
E.M. Nicholson and J.M. Scholtz, Biochemistry {\bf 35}, 11369 (1996), and
references therein.

\bibitem{low temp} Only a few proteins have been observed to denaturate
at normal conditions because their denaturation transition falls below
$0 ^o$C. Anyway, either raising the denaturation temperature by adding 
denaturants to
water, or by avoiding the freezing of water by supercooling or by applying
a pressure, cold denaturation has been observed in recent times
for many proteins~\cite{biochimica}.

\bibitem{Privalov90} P.L. Privalov, CRC Crit. Rev. Biochem. Mol. Biol. {\bf 25},
181 (1990).

\bibitem{PG88} P.L. Privalov and S.J. Gill, Adv. Protein Chem. {\bf 39}, 191 
(1988).

\bibitem{ISPHBT99} E.G. Isaacs, A. Shukla, P.M. Platzman, 
D.R. Hamann, B. Barbiellini and C.A. Tulk, Phys. Rev. Lett. {\bf 82}, 600
(1999); G.M. Gale, G. Gallot, F. Hache, N.Lascoux, S. Bratos, J.-Cl. Leicknam,
Phys. Rev. Lett. {\bf 82}, 1068 (1999).

\bibitem{HJSZ97} A. Hansen, M.H. Jensen, K. Sneppen and G. Zocchi,
Eur. Phys. J. B {\bf 6}, 157 (1998).

\bibitem{CCGP98} S. Caracciolo, M.S. Causo, P. Grassberger and 
A. Pelissetto, cond-mat/9812267.

\end{thebibliography}
\end{document}